\documentclass[12pt, draftclsnofoot, onecolumn]{IEEEtran}
\usepackage{cite}
\usepackage{graphicx,amsmath,amssymb}
\usepackage{subfigure}
\usepackage{citesort}
\usepackage{fancyhdr}
\usepackage{mdwmath}
\usepackage{mdwtab}
\usepackage{balance}
\usepackage{xcolor}
\usepackage{bm}

\usepackage{algorithm}
\usepackage{algorithmic}
\usepackage{multirow}
\usepackage{flafter}
\usepackage{comment}
\usepackage{amsthm}

\usepackage{multicol,lipsum}

\usepackage{float}
\floatstyle{plaintop}
\restylefloat{table}

\newtheorem{remark}{Remark}
\newtheorem{theorem}{Theorem}

\newtheorem{corollary}{\textbf{Corollary}}

\begin{document}


\title{Active Simultaneously Transmitting and Reflecting (STAR)-RISs: Modelling and Analysis}

%

\author{Jiaqi\ Xu, Jiakuo Zuo, Joey Tianyi Zhou, and Yuanwei\ Liu.

\thanks{J. Xu and Y. Liu are with the School of Electronic Engineering and Computer Science, Queen Mary University of London, London E1 4NS, UK. (email:\{jiaqi.xu, yuanwei.liu\}@qmul.ac.uk).}
\thanks{J. Zuo is with School of Internet of Things, Nanjing University of Posts and Telecommunications, Nanjing, China. (email: zuojiakuo@njupt.edu.cn)}
\thanks{J.T. Zhou is with Centre for Frontier AI Research (CFAR), A*STAR, 138632, Singapore. (e-mail: zhouty@cfar.a-star.edu.sg).}
}
\maketitle
\begin{abstract}
A hardware model for active simultaneously transmitting and reflecting reconfigurable intelligent surfaces (STAR-RISs) is proposed consisting of reflection-type amplifiers.
The amplitude gains of the STAR element are derived for both coupled and independent phase-shift scenarios.
Based on the proposed hardware model, an active STAR-RIS-aided two-user downlink communication system is investigated. Closed-form expressions are obtained for the outage probabilities of both the coupled and independent phase-shift scenarios. To obtain further insights, scaling laws and diversity orders are derived for both users.
Analytical results confirm that active STAR-RIS achieves the same diversity orders as passive ones while their scaling laws are different.
It is proved that average received SNRs scale with $M$ and $M^2$ for active and passive STAR-RISs, respectively.
Numerical results show that active STAR-RISs outperform passive STAR-RISs in terms of outage probability especially when the number of elements is small.

\end{abstract}

\begin{IEEEkeywords} 
Active RIS, hardware modelling, simultaneous transmitting and reflecting (STAR).
\end{IEEEkeywords}

\section{Introduction}
Over the past few years, simultaneous transmitting and reflecting reconfigurable intelligent surfaces (STAR-RISs)~\cite{liu_star,xu_star} have emerged as promising wireless technologies owing to their ability to favorably reconfigure the wireless communication environment on their both sides. 
STAR-RISs are composed of large amount of low-cost element whose transmission and reflection (T\&R) coefficients can be manually controlled. For passive-lossless designs, depending on the electric and magnetic properties of the STAR-RIS, the elements can have independent T\&R phase-shift~\cite{mu_star} or coupled T\&R phase-shift~\cite{xu_correlated}. 
Although STAR-RISs provide a wider coverage and better flexibility than conventional RISs, both RISs and STAR-RISs suffer from the ``double-fading'' effect where the small-scale fading of the base station (BS)-RIS link and the RIS-receiver link are multiplied~\cite{liu2020reconfigurable}. Even for STAR-RIS-aided \textit{virtual} line-of-sight (LoS) transmission, the received power falls off like $(d_1d_2)^{-2}$, where $d_1$ and $d_2$ are the path lengths of the two links mentioned above~\cite{xu_star}. As a result, a huge number of passive elements is required to achieve any practical performance gain. At the same time, the large surface area of STAR-RISs imposes challenges in terms of its deployment and maintenance.
To combat this critical problem, recently, the concept of active RISs was proposed where amplifiers are integrated the active elements.
In \cite{activeRIS}, the concept of active RISs was proposed and the joint optimization for reflecting phase-shift and receive beamforming was studied. In~\cite{zhang2021active}, a verified signal model was proposed for the active RISs.
However, there is currently a lack of research contribution on the modelling and performance analysis of active STAR-RISs. As a result, it is not clear how amplifiers can be integrated into STAR element for enhancing the amplitude of both transmitted and reflected signals. Furthermore, how T\&R phase-shift for active STAR elements are correlated remains unknown.

To answer these critical questions, this letter proposes a hardware model for active STAR-RISs with both coupled and independent T\&R phase-shift. Specifically, we show how amplification of the T\&R coefficients can be achieved by exploiting reflection-type amplifiers and quadrature hybrid couplers. Based on the proposed hardware model for active STAR-RISs, we analyze the outage probabilities for a two-user active STAR-RIS-aided communication system. Specifically, we obtain scaling law for the signal-to-noise ratio (SNR) and derive diversity orders by exploiting asymptotic analysis. Analytical results reveal that full diversity order can be achieved by both users for active STAR-RISs with independent phase-shift. We also show that unlike the quadrature scaling law for the passive STAR-RISs, the received SNR scales linearly with the number of active elements. 

In terms of energy consumption, a STAR-RIS can be passive lossy, passive lossless or active.
Under dual-sided incidence~\cite{dual}, each STAR element has four T\&R coefficients, i.e., $\Tilde{R}^A_m$, $\Tilde{R}^B_m$, $\Tilde{T}^{AB}_m$, and $\Tilde{R}^{BA}_m$. These four coefficients transform the incident signals to the output signals of each STAR element. The relation between the incident and output signals is given by:
\vspace{-0.1in}
\begin{equation}\label{tr_def}
    \begin{pmatrix}
y^A_m\\
y^B_m
\end{pmatrix} =
    \begin{pmatrix}
\Tilde{R}^A_m & \Tilde{T}^{AB}_m\\
\Tilde{T}^{BA}_m & \Tilde{R}^B_m
\end{pmatrix}
\begin{pmatrix}
s^A_m\\
s^B_m
\end{pmatrix},
\end{equation}
or equivalently, $\mathbf{y}_m = \mathbf{\Xi}_m \cdot \mathbf{s}_m$, where $\mathbf{\Xi}_m$ is the T\&R matrix of the $m$th STAR element. According to microwave network analysis, a STAR element is passive lossless if $\mathbf{\Xi}_m$ is unitary, i.e., $\mathbf{\Xi}_m^H\mathbf{\Xi}_m = \mathbf{I}_2$, where $\mathbf{I}_2$ denotes the two-by-two identity matrix. For passive lossy STAR element, $\mathbf{\Xi}_m^H\mathbf{\Xi}_m \prec \mathbf{I}_2$, and for active STAR element, $\mathbf{\Xi}_m^H\mathbf{\Xi}_m \succ \mathbf{I}_2$~\cite{pozar2011microwave}.
In terms of reciprocity, a STAR-RIS can be either reciprocal or nonreciprocal. A STAR element is reciprocal if the T\&R matrix $\mathbf{\Xi}_m$ is symmetrical, i.e., $\Tilde{T}^{AB}_m = \Tilde{T}^{BA}_m$. For nonreciprocal STAR matrix, we have $\Tilde{T}^{AB}_m \neq \Tilde{T}^{BA}_m$. 
In the letter, we focus on hardware models for active STAR-RISs with reciprocal elements where both independent and coupled phase-shift scenarios are considered.

\begin{figure}[b!]
    \begin{center}
        \includegraphics[scale=0.4]{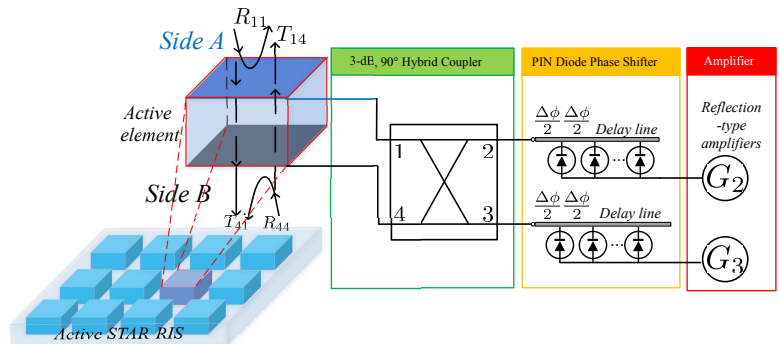}
        \caption{Hardware structure of an active STAR element.}
        \label{hybrid}
    \end{center}
\end{figure}

\section{Hardware and System Models}

As illustrated in Fig.~\ref{hybrid}, the proposed active STAR-RIS element consists of three major components: a quadrature hybrid~\cite{pozar2011microwave}, positive-intrinsic–negative (PIN) diode phase-shifters, and reflection-type amplifiers~\cite{amp}.
Upon receiving the wireless signal, a 3-dB, $90^\circ$ hybrid coupler with all four ports matched is employed to couple signals incident from both sides and minimize power loss. Then, PIN diode switches and phase-shifters are connected to port 2 and port 3 of the hybrid coupler to impose adjustable phase delay to the signal. Finally, the delayed signals are fed to reflection-type amplifiers. 
%
Depending on the configurations of the phase-shifters and the gains of the amplifiers, the active STAR element structure has different overall transmission and reflection behavior.
In the following, we formulate the T\&R coefficient of the active STAR element in terms of the phase-shift values of the PIN diode phase-shifters and the gains of the reflective amplifiers.

\subsection{Hardware Model for Active STAR-RISs}

\subsubsection{Active STAR Element with Coupled T\&R Phase-Shift}
If only one delay line and one reflection-type amplifier are available for each STAR element, the resultant element will exhibit coupled transmission and reflection coefficients. As illustrated in Fig.~\ref{hybrid}, the delay line and amplifier can be connected to either port 2 or port 3 of the hybrid coupler. Without loss of generality, suppose that port 2 is connected to the reflection-type amplifier with gain $G_2$ and port 3 is grounded i.e., $V_2^+ = \Tilde{G}_2V_2^-$ and $V_3^+ = 0$. Exploiting the scattering matrix of the quadrature hybrid, the output and input signal voltages have the following relation:
\begin{equation}\label{mat}
    \begin{pmatrix}
V_1^-\\
V_2^-\\
V_3^-\\
V_4^-
\end{pmatrix} =
    -\frac{1}{\sqrt{2}}\begin{pmatrix}
0 &j &1 &0\\
j &0 &0 &1\\
1 &0 &0 &j\\
0 &1 &j &0
\end{pmatrix}
    \begin{pmatrix}
V_1^+\\
V_2^+\\
V_3^+\\
V_4^+,
\end{pmatrix}
\end{equation}
where $V_2^+ = \Tilde{G}V_2^-$, $V_3^+=0$, and $\Tilde{G} =G_2\cdot e^{j\phi_2}$ is the complex amplitude response of the delay line and amplifier combined. 
Using the relation given in \eqref{mat}, the output signals through port 1 and port 4, i.e., the wireless signals exit from \textit{side A} and \textit{side B} of the STAR element can be derived as follows:
\begin{align}\label{v1}
    &V_1^- = -\frac{\Tilde{G}}{2}\cdot V_1^+ + \frac{j\Tilde{G}}{2}\cdot V_4^+\\ \label{v2}
    &V_4^- = \frac{j\Tilde{G}}{2}\cdot V_1^+ + \frac{\Tilde{G}}{2}\cdot V_4^+.
\end{align}
\begin{theorem}
For STAR elements with coupled T\&R phase-shift, the four T\&R coefficients of the STAR elements which are connected to an amplifier with gain $\Tilde{G}$ can be expressed as follows:

\begin{align}\label{TR}
    &R^A_m = \frac{V_1^-}{V_1^+}\Big|_{V_4^+=0}\!=\!-\Tilde{G}/2,\ \ R^B_m = \frac{V_4^-}{V_4^+}\Big|_{V_1^+=0}\!=\!\Tilde{G}/2 \\
    &T^{AB}_m = \frac{V_1^-}{V_4^+}\Big|_{V_1^+=0}\!=\!j\Tilde{G}/2,\ \ T^{BA}_m = \frac{V_4^-}{V_1^+}\Big|_{V_4^+=0}\!=\!j\Tilde{G}/2.
\end{align}
\begin{proof}
The proof is straightforward by exploiting the definitions of the T\&R coefficients and the results obtained in \eqref{v1} and \eqref{v2}.
\end{proof}
\end{theorem}

\begin{remark}
    It can be observed that for the case where one amplifier is employed per element, the T\&R coefficients are highly coupled.
    Firstly, since the overall hardware structure is reciprocal, we have $T^{AB}_m = T^{BA}_m$. Next, consider the T\&R coefficients which are involved in the single-side incidence from \textit{side A}, i.e., $R^{A}_m$ and $T^{BA}_m$, we have $(\phi^R_m - \phi^T_m)_{A} = \angle{R^A_m} - \angle{T^{BA}_m} = \pi/2$. Similarly, for \textit{side B}, we have $(\phi^R_m - \phi^T_m)_{B} = \angle{R^B_m} - \angle{T^{AB}_m} = -\pi/2$. Note that the phase difference between the T\&R coefficients for the two sides will be exchanged if the amplifier is connected to port 3. This indicates that the active coupled phase-shift STAR-RIS follows the same T\&R phase-shift correlation as the passive-lossless STAR-RISs, i.e., $\phi^R_m - \phi^T_m = \pm \pi/2$~\cite{xu_correlated}. Finally, the amplitude of all four T\&R coefficient are identical and they are determined by the gain of the amplifier $G_2$.
\end{remark}

\subsubsection{Active STAR Element with Independent T\&R Phase-Shift}
If two amplifiers are available for each STAR element, the resultant active STAR element has the ability to independently adjust the amplitude and phase-shift of the T\&R coefficients. As illustrated in Fig.~\ref{hybrid}, the gains of the the two amplifiers are $G_2$ and $G_3$. We further denote $\Tilde{G}_2 = G_2e^{j\phi_2}$ and $\Tilde{G}_3 = G_3e^{j\phi_3}$ as the combined gain and phase-shift delay of port 2 and port 3, respectively. 
By substituting $V_2^+ = \Tilde{G}_2V_2^-$ and $V_3^+ =\Tilde{G}_3V_3^- $ into \eqref{mat}, the output signals at port 1 and port 4 are given by:
\begin{align}\label{v1'}
    &V_1^- = \frac{\Tilde{G}_3-\Tilde{G}_2}{2}\cdot V_1^+ + \frac{j(\Tilde{G}_2+\Tilde{G}_3)}{2}\cdot V_4^+\\ \label{v2'}
    &V_4^- = \frac{j(\Tilde{G}_2+\Tilde{G}_3)}{2}\cdot V_1^+ +  \frac{\Tilde{G}_2-\Tilde{G}_3}{2}\cdot V_4^+.
\end{align}
\begin{theorem}
For STAR elements with independent active T\&R phase-shift, the four T\&R coefficients of the STAR elements which are connected to amplifiers with gain $\Tilde{G}_2$ and $\Tilde{G}_3$ can be expressed as follows:
\begin{align}\label{TR'}
    &R^A_m = -R^B_m = (\Tilde{G}_3-\Tilde{G}_2)/2 \\
    &T^{AB}_m = T^{BA}_m = j(\Tilde{G}_2 + \Tilde{G}_3)/2.
\end{align}
\begin{proof}
The proof is straightforward by exploiting the results obtained in \eqref{v1'} and \eqref{v2'}.
\end{proof}
\end{theorem}
\begin{remark}
    It can be observed that since the amplitudes and phases of both $G_2$ and $G_3$ can be adjusted by tuning the phase-shift delay lines and amplifiers, the amplitudes and phases the T\&R coefficients can be independently configured. 
    Moreover, the transmitting-only and reflecting-only modes can be achieved by letting $\Tilde{G}_2 = \Tilde{G}_3$ and $\Tilde{G}_2 = -\Tilde{G}_3$, respectively.
    However, due to the reciprocal design, the phase-shift and amplitudes of the transmission coefficients in the two directions remain identical, i.e., $T^{AB}_m = T^{BA}_m$.
\end{remark}

\begin{figure*}[t!]
\centering
\subfigure[Coupled phase-shift, active]{\label{n0}
\includegraphics[width= 1.5in]{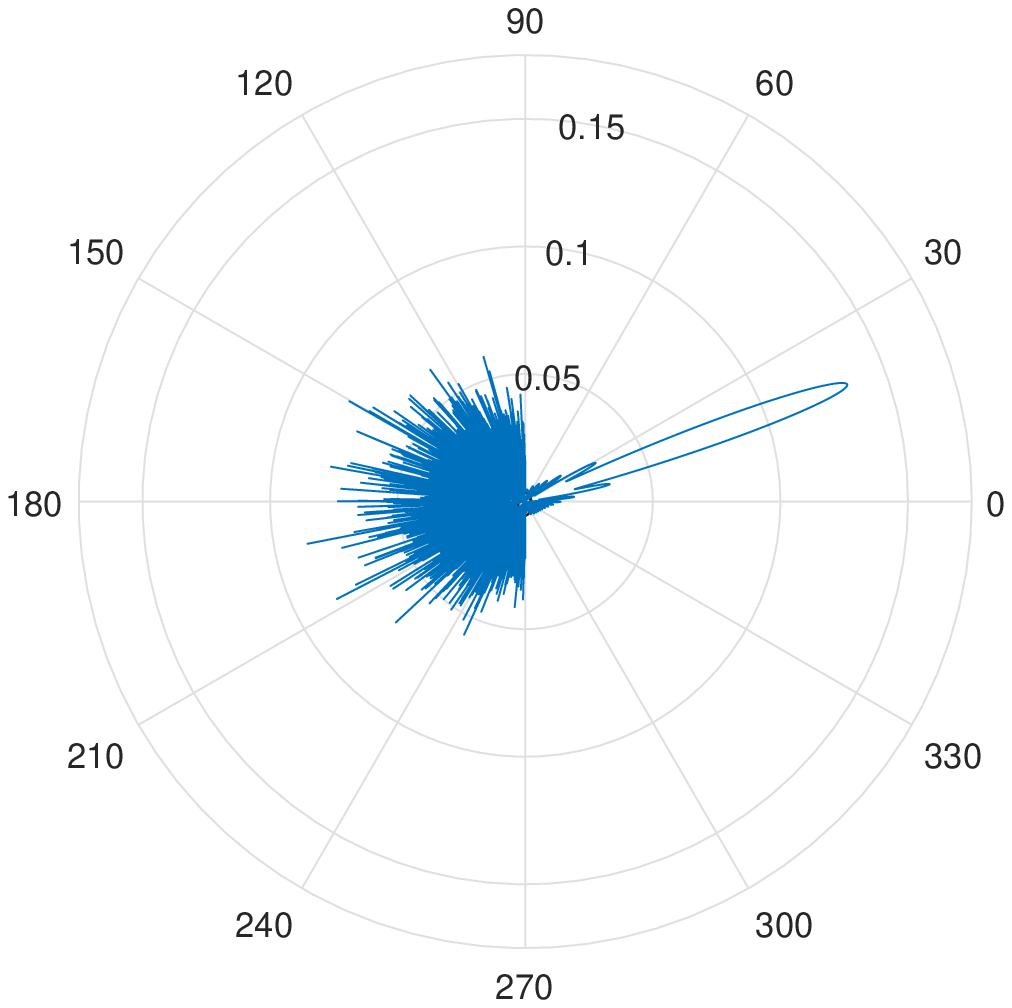}}
\subfigure[Independent phase-shift, active]{\label{na}
\includegraphics[width= 1.5in]{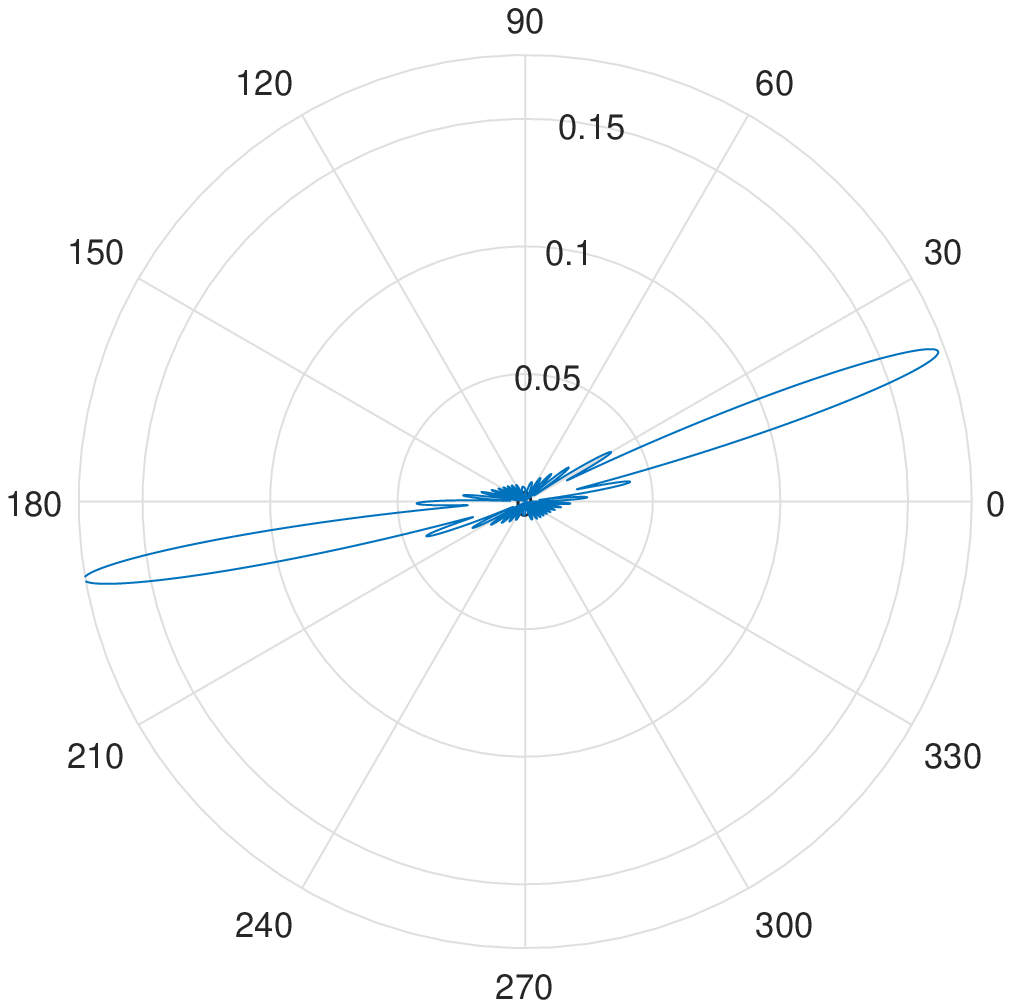}}
\subfigure[Passive-lossless STAR-RIS]{\label{nb}
\includegraphics[width= 1.5in]{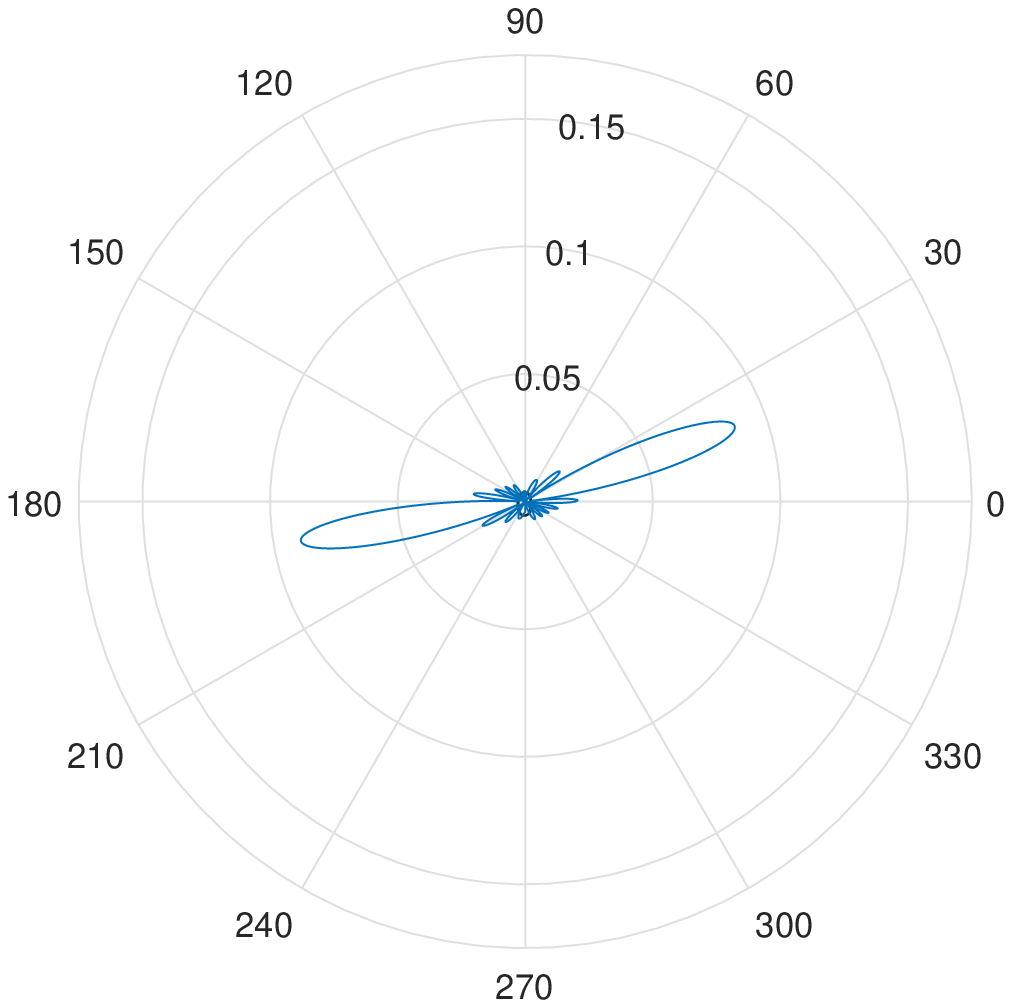}}
\caption{Simulated radiation pattern of active and passive STAR-RISs.}\label{nice}
\end{figure*}
\vspace{-0.2in}
\subsection{System Model and Radiation Pattern}
We investigate a two-user downlink communication system where single-antenna base station (BS) is located on \textit{side A} and the two users are locate on different sides of the STAR-RIS.
The direct link from BS to both users are assumed to be severely blocked. As a result, an active STAR-RIS is deployed to recover the service for the two users.
Let $g_m$ denote the channel between the BS and the $m$th active STAR element, $h^{\chi}_m$ denote the channel between the $m$th active STAR element and the user $\chi$ where $\chi \in \{A,B\}$ denotes the user on \textit{side A} or \textit{side B}. All channel are assumed to exhibit Rician fading, i.e., $|h^\chi_m|\sim\mathcal{R}(K^\chi_h,\Omega^\chi_h)$ and $|g_m|\sim\mathcal{R}(K_g,\Omega_g)$, where $K^\chi_h$ and $K_g$ are the shape parameters, $\Omega^\chi_h$ and $\Omega_g$ are the scale parameters of the corresponding Rician channels. 
Considering the noise at the active STAR element and at each users, the received signal for user $\chi$ in the downlink scenario is given by:
\begin{equation}
    y^\chi = \sum_{m=1}^M h^\chi_m G^\chi_m e^{j\phi^\chi_m}  (g_m s_m + v_m) + n^\chi
\end{equation}
where $G^\chi_m$ and $\phi^\chi_m$ are the amplitude and phase-shift of the T\&R coefficient of the $m$th element corresponding to user $\chi$, respectively. $v_m$ is the noise caused by the $m$th active STAR element, $n^\chi$ is the noise caused by user $\chi$. 
We further assume that the channel state information is known at the active STAR-RIS. As a result, cophase condition can be adopted to align the phase-shift of each active STAR signals at both users.

In Fig.~\ref{nice}, we illustrate the angular patterns of the active STAR-RIS with coupled and independent T\&R phase-shift and the passive-lossless STAR-RIS as a baseline. In all scenarios, we consider a STAR-RIS with 18$\times$18 elements.
We assume that user $A$ and user $B$ are located at angular positions of $20^\circ$ and $190^\circ$ with equal distances of $10$ m from the STAR-RIS, respectively. 
As can be seen from Fig.~\ref{n0}, the coupled phase-shift active STAR-RIS forms a narrow beam towards user $A$. However, for the transmission space (the second and third quadrants), there are no observable beams. This is because due to the phase-shift coupling, only the reflection phase-shift follows the cophase condition for user $A$. The radiation pattern for independent active STAR-RIS is shown in Fig.~\ref{na}. Compared with the coupled phase-shift scenario, the signal strength is focused for both user $A$ and $B$. Moreover, based on the proposed hardware models, the each independent active STAR element employs one more amplifiers compared to the coupled phase-shift STAR element. As a result, the independent phase-shift STAR-RIS achieve larger amplitudes for both reflected and transmitted beams. In comparison, the passive-lossless STAR-RIS shown in Fig.~\ref{nb} has the smallest reflection and transmission beam amplitudes.

\section{Performance Analysis}
To evaluate the performance of the proposed active STAR-RIS, in this section, we study a active STAR-RIS-aided wireless communication system serving two users. Based on the proposed active STAR-RIS hardware model, the outage probabilities and their asymptotic behaviours are analyzed.
For both coupled and independent phase-shift scenario, the received SNR of user $\chi$ is given as follows:
\begin{align}\label{ga}
    \gamma^\chi &= \frac{p|\sum_{m=1}^M h^\chi_m G^\chi_m e^{j\phi^\chi_m} g_m|^2}{|\sum_{m=1}^Mh^\chi_mG^\chi_m e^{j\phi^\chi_m}|^2\sigma^2_v + \sigma^2_\chi}
\end{align}
where $P$ is the transmit power of the BS, $\sigma_v^2$ is the variance of $v_m$, $\sigma^2_{\chi}$ is the variance of $n^{\chi}$.  

\subsection{Scaling Laws}
The following theorems give the scaling laws for the SNRs of both users. We investigate the received SNR of both users in the high transmit SNR limit, i.e., $p/\sigma^2_\chi \to \infty$. For active STAR-RISs with coupled T\&R phase-shift, assume that the phase-shift reflection coefficients of the active element are configured to align the signals for user $A$. Due to the coupling between $R^A_m$ and $T^{BA}_m$, the phase-shift of the transmission coefficients are not optimized for user $B$.

\begin{theorem}\label{t_couple}
    For active STAR-RISs with coupled T\&R phase-shift, the scaling laws for the received SNRs are given as follows:
    \begin{align}\label{bga}
        \overline{\gamma^A} = \frac{p|MG_{m}\overline{h^A_m}\overline{g}_m|^2}{M\pi G_m^2\overline{(h^A_m)^2}\sigma^2_v/4 + \sigma^2_A} \to \frac{p}{\sigma_v^2}\frac{4\overline{h^A_m}\overline{g}_m}{\pi\overline{(h^A_m)^2}} \cdot M\\ \label{17}
        \overline{\gamma^B} = \frac{pMG_{m}^2\pi \overline{(h^B_mg_m)^2}/4}{M\pi G_{m}^2\overline{(h^B_m)^2}\sigma^2_v/4 + \sigma^2_B} \to \frac{p}{\sigma_v^2}\frac{\overline{(h^B_mg_m)^2}}{\overline{(h^B_m)^2}},
    \end{align}
    where $\overline{x}$ denotes the expected value of a random variable $x$, $G_{m}$ denotes the maximum gain of an active STAR element.
    \begin{proof}
        For user A, the reflection coefficients are configured to align the phase-shifts of the cascaded channels $h^A_mg_m$. Thus, we have $\sum h^A_mg_m R^A_m = \sum|h^A_m||g_m||R^A_m|$. Next, since $h^A_m$ and $g_m$ are independent Rician variables, the phases of each terms of $h^A_mR^A_m$ can be regarded as randomly distributed. According to the law of large numbers, the expectation value of the sum in the denominator can be calculated as $\mathbb{E}[|\sum h^A_mR^A_m|]=\frac{1}{2}\sqrt{M\pi\overline{(h^A_m)^2}|R^A_m|^2}$. For user B, its channel experience random phase alignment due to the T\&R phase-shift coupling. Both summations in \eqref{ga} for $\chi = B$ (user $B$) have randomly distributed phase-shifts and their expectation value can be obtained in similar fashion using the law of large numbers.
    \end{proof}
\end{theorem}
\begin{remark}
    For the coupled T\&R phase-shift scenario, the SNRs for user $A$ scales with $M$ and the SNR of user $B$ is upper bounded by a constant value. Recall that for passive STAR-RISs, the SNRs for users scales with $M^2$. Although passive STAR-RISs have better performance asymptotically, active STAR-RISs are able to outperform passive ones before reaching the large $M$ limit, i.e., with a practical number of elements. More importantly, the gain $G_m$ introduced by the active elements can significantly increase users SNRs for the cases where $M$ is small, as can be observed in \eqref{bga}.
\end{remark}

\begin{theorem}
    For active STAR-RISs with independent T\&R phase-shift, the scaling laws for the two users can be calculated as follows:
    \begin{align}
        \overline{\gamma^\chi} = \frac{p|MG^\chi_{m}\overline{h^\chi_m}\overline{g}_m|^2}{M\pi (G^\chi_m)^2\overline{(h^\chi_m)^2}\sigma^2_v/4 \!+\! \sigma^2_\chi} \to \frac{p}{\sigma_v^2}\frac{4\overline{h^\chi_m}\overline{g}_m}{\pi\overline{(h^\chi_m)^2}} \cdot M
    \end{align}
    where $G^A_m=|(\Tilde{G}_3-\Tilde{G}_2)/2|$ is the magnitude of the reflection coefficient of the independent phase-shift active STAR element and $G^B_m = |(\Tilde{G}_3+\Tilde{G}_2)/2|$ is the magnitude of the transmission coefficient.
    \begin{proof}
        The proof of this theorem follows a similar process as in Theorem~\ref{t_couple}.
    \end{proof}
\end{theorem}
\subsection{Outage Probability and Diversity Orders}
The outage probability of user $\chi$ is defined as follows:
\begin{equation}\label{out_def}
P_{out}^{\chi} = Pr\Big\{\gamma^\chi < \gamma_{target}^\chi  \Big\},
\end{equation}
where $\gamma^\chi$ is the received SNR of user $\chi$, which is given in \eqref{ga}. For simplicity, we further assume that the gains of the STAR elements are the same, i.e., $G^\chi_m = G^\chi,\ \forall m\in\{1,\cdots,M\}$. In the following, we investigate the outage probabilities of both users for both coupled and independent T\&R phase-shift scenarios in two theorems.

\begin{theorem}\label{t_pri}
For user $A$ under coupled phase-shift STAR-RIS or both users under independent phase-shift scenario, their asymptotic outage probabilities can be expressed as follows:
\begin{align}\label{out1}
P^R_{out}(\gamma^\chi_{\text{target}})=\Big[ \frac{4(K^\chi_h\!+\!1)(K_g\!+\!1)\sigma_\Sigma^2\gamma^\chi_{\text{target}}}{\Omega_h^\chi\Omega_ge^{K^\chi_h+K_g}G^\chi}\Big]^M \frac{p^{-M}}{(2M)!},
\end{align}
where $\sigma^2_\Sigma = M\Omega^\chi_hG^\chi\sigma^2_v + \sigma^2_\chi$ is the variance of the combined noise at the active STAR elements and at the receiver.
\begin{proof}
See Appendix~A.
\end{proof}
\end{theorem}

Based on the derived asymptotic outage probability, we investigate the diversity order of user $\chi$, which is defined as follows:
\begin{equation}\label{d_define}
d^{\chi} = -\lim_{p \to \infty}\frac{\log P^{\chi}_{out}(\gamma^\chi_{\text{target}})}{\log p}.
\end{equation}
\begin{corollary}
The diversity orders of user $A$ under coupled phase-shift STAR-RIS or both users under independent phase-shift scenario are given by:
\begin{equation}\label{d_pri}
    d^\chi_{\text{independent}} = d^A_{\text{coupled}} = M.
\end{equation}
\begin{proof}
Combining \eqref{out1} with \eqref{d_define}, it is straightforward to obtain \eqref{d_pri}.
\end{proof}
\end{corollary}

\begin{theorem}\label{t_2}
For user $B$ under coupled phase-shift scenario, its outage probabilities can be expressed as follows:
\begin{equation}\label{out2}
P^B_{out}(\gamma^B_{\text{target}})=1-\exp\{
-\frac{\gamma^B_{\text{target}}\sigma^2_\Sigma}{M\Omega^B_h\Omega_g(G^\chi)^2 p}
\},
\end{equation}
where $\sigma^2_\Sigma = M\Omega^B_hG^B\sigma^2_v + \sigma^2_B$ is the variance of the combined noise at the active STAR elements and at receiver $B$.
\begin{proof}
For user $B$ under coupled phase-shift scenario, the numerator of \eqref{ga} undergo random phase-shift alignment. Thus, the overall channel of $|H^B| = |\sum_m^M h^\chi_mg_me^{j\phi^\chi_m}|$ follows a Rayleigh distribution, i.e., $f_{|H^B|}(x) = \frac{2x}{\Omega^B}e^{-x^2/\Omega^B}$, where $\Omega^B = M\Omega^B_h\Omega_g$~\cite{xu_correlated}. Then, the outage probability of user $B$ can be derived by exploiting its definition given in \eqref{out_def}.
\end{proof}
\end{theorem}

\begin{corollary}
The diversity order of user $B$ under coupled phase-shift STAR-RIS is given by:
\begin{equation}\label{d_2}
    d^B_{\text{coupled}} = 1.
\end{equation}
\begin{proof}
Combining \eqref{out2} with \eqref{d_define} and exploiting the fact that the Taylor expansion of $e^x = 1+x+o(x^2)$ near the origin, it is straightforward to obtain \eqref{d_pri}.
\end{proof}
\end{corollary}
\subsection{Summary}
In Table~I, we summarize the analytical insights obtained from this section. As shown in the table, the sum diversity order of the two users and the received SNR scaling laws are compared for active and passive STAR-RISs.
\begin{table}[!h]
\centering
\small
\label{tab:my-table1}
\begin{tabular}{|c|c|c|}
\hline
STAR-RISs           & Sum Diversity Order & Scaling Law \\ \hline
Active, Coupled     & $M+1$               &  $\propto M$ or bounded          \\ \hline
Active, Independent & $2M$                &         $\propto M$     \\ \hline
Passive Lossless    & $2M$                &         $\propto M^2$     \\ \hline
\end{tabular}
\caption{Summary of diversity order and scaling law for active and passive STAR-RISs}
\end{table}

\section{Numerical Results}\label{num}

In this section, simulation results are provided to investigate the performance of the active STAR-RIS-aided communication. For our simulations, we assume that the active STAR-RIS is a uniform planar array consisting of $M$ elements. The spacing between adjacent elements is half of the carrier wavelength. 
The noise power for both users is set to $\sigma^2_A = \sigma^2_B = -10$ dBm and the noise power of the active element is set to $\sigma^2_v = -20$ dBm. Moreover, the gain of the reflection-type amplifiers are set to $G_2=G_3 \approx 1.5$ dB.
All STAR-RIS-user channels are modeled as Rician fading channels with path loss exponents of $\alpha = 2.2$ and the Rician factor of $K=1.5$ dB.

\begin{figure}[t!]
    \begin{center}
        \includegraphics[scale=0.5]{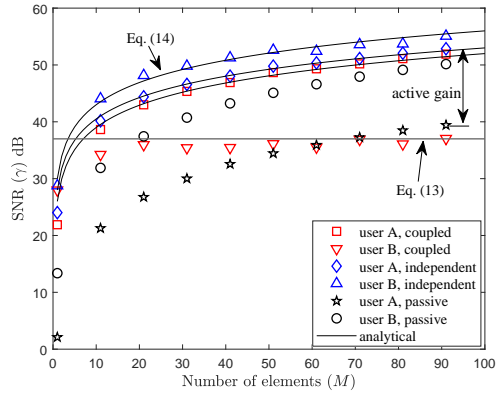}
        \caption{\textcolor{black}{Simulated and analytical results for SNR scaling laws of user $A$ and user $B$.}}
        \label{op2}
    \end{center}
\end{figure}

\begin{figure}[t!]
    \begin{center}
        \includegraphics[scale=0.5]{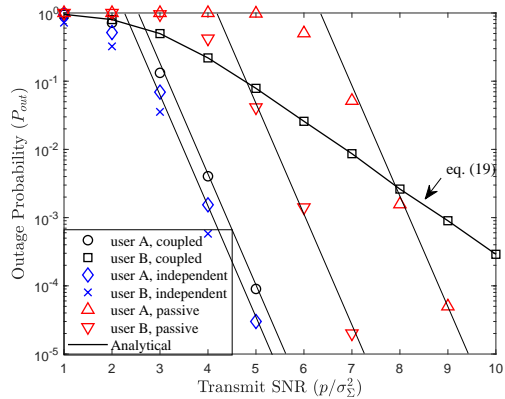}
        \caption{Simulated and analytical asymptotic results for outage probabilities of user $A$ and user $B$.}
        \label{fout1}
    \end{center}
\end{figure}

In Fig.~\ref{op2}, the received SNR is plotted against the number of STAR elements. 
As can be observed in the figure, the independent active STAR-RIS has the highest SNRs and outperforms both active STAR-RIS with coupled phase-shift and passive STAR-RIS. For the coupled phase-shift active STAR-RIS, the received SNR for user $A$ always outperforms the passive STAR-RIS. However, in terms of the received SNR for user $B$, the passive STAR-RIS outperform coupled phase-shift active STAR-RIS after $M > 70$. This is due to the fact that the received SNR of user $B$ is upper bounded as derive in \eqref{17}. Based on these observations, we can conclude that the perform gain of active STAR-RIS over passive ones is more significant when there is size restriction or the number of elements is small.

In Fig.~\ref{fout1}, we investigate the outage probabilities for both users for the case of coupled phase-shift active STAR-RIS, independent phase-shift active STAR-RIS, and passive-lossless STAR-RIS. As can be observed, the independent phase-shift active STAR-RIS achieves the lowest outage probabilities for both users. For the case of coupled phase-shift STAR-RIS, user $A$ has a lower outage compared to the passive baseline. However, user $B$ has a smaller diversity order as derived in \eqref{d_2}. As a result, the passive STAR-RIS outperform the coupled phase-shift STAR-RIS when transmit SNR is high. Based on these observations, we can conclude that the perform gain of active STAR-RIS over passive ones is more significant when the transmit SNR is lower.

\section{Conclusions}
In this letter, we provided a hardware model for active STAR-RISs with both coupled and independent phase-shift.
The amplitudes and phase-shifts of the T\&R coefficients of the active STAR element were derived using the gain of the reflection-type amplifiers.
Based on the proposed hardware model, we studied the performance of an active STAR-RIS-aided two user wireless communication system.
We derived expressions for the asymptotic received SNRs and outage probabilities of both users for the case of coupled phase-shift and independent phase-shift. It was proved that both user can achieve full diversity order under independent phase-shift active STAR-RIS. Finally, numerical results demonstrated the performance gain of active STAR-RIS over passive ones. It was also revealed that the performance gain is more significant when the number of STAR element is small and the transmit SNR is low.

\begin{appendices}

\renewcommand{\theequation}{A.\arabic{equation}}
\setcounter{equation}{0}
\section{Proof of \textbf{Theorem~\ref{t_pri}}}\label{ap_b}
In the high transmit SNR limit, we approximate the overall received noise, i.e., the denominator of \eqref{ga} with its variance $\sigma^2_\Sigma$. Since the phase of each term in the summation $\sum_m^M h^\chi_mG^\chi_me^{j\phi^\chi_m}$ is randomly aligned, we have $\sigma^2_\Sigma = M\Omega^\chi_hG^\chi\sigma^2_v + \sigma^2_\chi$.
Next, we consider the nominator of \eqref{ga},
according to the system model, both $|g_m|$ and $|h^\chi_m|$ follow Rician distributions. After the phase alignment of active STAR element, we only need to consider the distribution of $H^\chi =\sum_m^M |g_m||h^\chi_m|$. Each term in the summation is the product of two independent Rician random variables whose probability distribution function (PDF) is given in~\cite{simon2002probability}. 
We approximate the PDF of $|g_m||h^\chi_m|$ near the origin using a Taylor series expansion:
\begin{equation}\label{ap_pro_asymp}
 f_{|h^\chi_m||g_m|}(x) = \frac{4(K^\chi_h+1)(K_g+1)}{\Omega^\chi_h\Omega_ge^{(K^\chi_h+K_g)}} \cdot x + o(x),
\end{equation}
where $o(\cdot)$ is the little-o notation and $o(f(x))$ denotes a function which is asymptotically smaller than $f(x)$. 
Then Laplace transform of $|H^\chi|$ can be calculated as follows:

\begin{align}\label{l_H_o}
\begin{split}
    \mathcal{L}\{f_{|H^\chi|}(x)\}(t) &= \Big( \mathcal{L}\{f_{|h^\chi_m||g_m|}(x)\}(t)\Big)^M \\&= 
    \Big[\frac{4(K^\chi_h+1)(K_g+1)}{\Omega^\chi_h\Omega_ge^{(K^\chi_h+K_g)}}\Big]^M
    t^{-2M} +o(t^{-2M}).
    \end{split}
\end{align}
Finally, the PDF of $|H^\chi|$ can be obtained by performing the inverse Laplace transform of \eqref{l_H_o}. The outage probability can therefore be derived using the obtained PDF.
\end{appendices}

\bibliographystyle{IEEEtran}
\bibliography{mybib}

\end{document}